\documentclass[runningheads]{llncs}
\usepackage{graphicx}
\usepackage{amsmath}
\begin{document}
\title{The (de)biasing effect of GAN-based 
augmentation methods on skin lesion images}
%
%
\author{Agnieszka Mikołajczyk 
\inst{1}\orcidID{0000-0002-8003-6243} \and
Sylwia Majchrowska\inst{2,3}\orcidID{0000-0001-7576-7167} 
\and
Sandra Carrasco Limeros \inst{2,3}\orcidID{0000-0002-4782-7166}} 
\authorrunning{A. Mikołajczyk et al.}
%
\institute{Gdańsk University of Technology, Gabriela Narutowicza 11/12, 80-233 Gdańsk, Poland \\
\email{agnieszka.mikolajczyk@pg.edu.pl}\\
\and
Sahlgrenska University Hospital, Blå stråket 5, 413 45 Göteborg, Sweden \\
\and
AI Sweden, Lindholmspiren 3-5, 402 78 Göteborg, Sweden \\
\email{\{sylwia.majchrowska, sandra.carrasco\}@ai.se}}
\maketitle              
\begin{abstract}

New medical datasets are now more open to the public, allowing for better and more extensive research. Although prepared with the utmost care, new datasets might still be a source of spurious correlations that affect the learning process. Moreover, data collections are usually not large enough and are often unbalanced. One approach to alleviate the data imbalance is using data augmentation with Generative Adversarial Networks (GANs) to extend the dataset with high-quality images. GANs are usually trained on the same biased datasets as the target data, resulting in more biased instances. This work explored unconditional and conditional GANs to compare their bias inheritance and how the synthetic data influenced the models. We provided extensive manual data annotation of possibly biasing artifacts on the well-known ISIC dataset with skin lesions.
In addition, we examined classification models trained on both real and synthetic data with counterfactual bias explanations. Our experiments showed that GANs inherited biases and sometimes even amplified them, leading to even stronger spurious correlations. Manual data annotation and synthetic images are publicly available for reproducible scientific research.

\end{abstract}




\section{Introduction}

Deep learning-based approaches need a large amount of annotated data to perform well. High-quality images can be easily generated using publicly available pretrained Generative Adversarial Networks (GANs). It seems especially useful in medical applications like skin lesion classification, detection of lung cancer nodules, or even brain tumor segmentation, where balanced data is a definite must-have.

However, if GAN's training set is biased, augmentation might backfire instead of helping. Bias is often defined as \textit{a systematic error from erroneous assumptions in the learning algorithm}~\cite{mehrabi2021survey}. In this work, we focused primarily on bias in data and models. With the term 'bias in data,' we referred to four common data biases in machine learning (ML): \textit{observer bias} which might appear when annotators use personal opinion to label data~\cite{mahtani2018catalogue}; \textit{sampling bias} when not all samples have the same sampling probability~\cite{mehrabi2021survey}; \textit{data handling bias} when the way of handling the data distort the classifier's output; and \textit{instrument bias} meaning imperfections in the instrument or method used to collect the data~\cite{he2012bias}. By 'bias in models', we referred to the broad term of the algorithmic bias~\cite{baeza2018bias}. Some sources define an algorithmic bias as amplifying existing inequities in, e.g., socioeconomic status, race, or ethnic background by an algorithm~\cite{panch2019artificial}.

The problem of bias amplification is often mentioned e.g. in recommending engines~\cite{lloyd2018bias}, word embeddings~\cite{bolukbasi2016man}, or any other discriminate model~\cite{mayson2018bias}. This leads to the question: if these models can amplify biases, does GANs do it too? If it does, how strongly GAN-augmented data affects the models?

Hence, to answer those questions, we studied the influence of data augmentation with unconditional and conditional GANs in terms of possible bias amplification. We analyzed with counterfactual bias insertion (CBI) GAN's ability to reproduce artifacts observed in a dataset, such as frames, pen markings, and hairs. 
In addition, we evaluated GANs in terms of fidelity, diversity, speed of training, and performance of classifiers trained on mixed real and synthetic data. 

Our contributions are the following. Firstly, we performed the extensive research on the (de)biasing effect of using GAN-based data augmentation. Secondly, we introduced the dataset with manual annotations of biasing artifacts in six thousands synthetic and real skin lesion images, which can serve as a benchmark for further studies. Finally, we showed that the most represented biases in the real data are enhanced by the generative network whereas the least represented artifacts are reduced.

\section{Related works}
Previous studies have showed that skin lesion datasets are not free from bias. Winkler et al.~\cite{winkler2019association} proved that common artifacts like surgical pen markings are strongly correlated with the skin lesion type, influencing the model. Bissotto et al.~\cite{bissoto2019constructing} presented that certain artifacts in skin lesion datasets affect classification models so strongly that they achieve relatively good results even when the lesion is removed from the image. 
Using global explainability and CBI methods, Mikołajczyk et al.~\cite{mikolajczyk2020towards} examined how strongly artifacts can influence the training. The result showed that the model is strongly biased towards black frame artifacts, as inserting one into the image often leads to significant prediction shifts. Bevan et al.~\cite{bevan2022detecting} presented that skin tone is also a biasing factor that can influence the models. Considering the literature review, the most commonly mentioned artifacts are hair, rulers, frames, and others like gel bubbles or surgical pen markings. In the paper we examine and annotate those artifacts in Section~\ref{sec:data_stats}.

Some works on measuring bias suggested simply comparing the performance metrics on biased and unbiased dataset~\cite{park2018reducing}. But in a real-world scenario, it is usually not possible to access an unbiased dataset. Such an approach would require removing all biases before training. In the case of skin lesions, removing artifacts like black frames, surgical pen markings, and even hair is very difficult, especially when these artifacts are on top of the lesions. 
A CBI is a contrasting approach, where one needs to insert the bias instead~\cite{mikolajczyk2020towards}. CBI introduced a set of metrics that can be used to evaluate prediction shift after adding the bias to the input data: mean and median prediction shift and a number of $switched$ predictions. Higher rates mean a higher risk of giving biased predictions. As those numbers do not indicate the accuracy or correctness of the predicted category, it is worth measuring the $F_1$ score, recall, or other performance metrics to observe if the accuracy is lower on the dataset with inserted bias. 

The problem of instrument bias in melanoma classification for the ISIC2020 dataset was addressed before using different debiasing techniques for artifact bias removal~\cite{Bevan2021SkinDU}. However, the authors mitigated only two selected biases: the surgical marking and ruler. They investigated the generalization capabilities of the bias removal approaches across different CNN architectures and human diagnosis. On average, EfficientNet-B3,  ResNet-101, ResNeXt-101, DenseNet, and Inception-v3 models reached better accuracy ($AUC \approx  0.88$) than experienced dermatologists, performing similarly amongst themselves. In these studies, artificial data was not utilized to augment real data.

 
The generation of synthetic data not only increases the amount of data and balances the dataset but also serves as an anonymization technique that facilitates its exchange between different institutions as a proxy dataset~\cite{bib:ganrew}.
Despite many attempts to generate artificial samples of skin images, the evaluation methods for the generated data's quality, diversity, and authenticity are still unclear. In some works~\cite{bib:ISICCls22}, researchers point out the inadequacies of commonly used open datasets, such us data imbalance, bias or unwanted noise and artifacts. As GANs are learning the distributions of all provided images, they might as well learn and generate those unwanted features. 
\begin{figure}[t] \label{fig:pipeline}
\includegraphics[width=\textwidth]{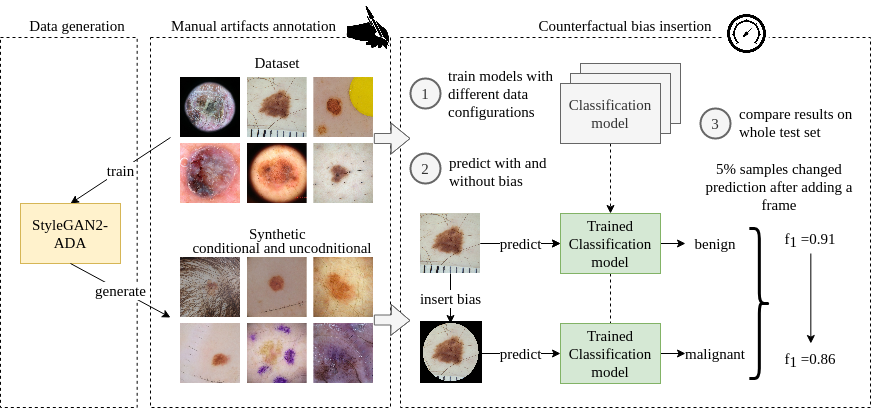}
\centering
\caption{The procedure behind (de)biasing effect of using GAN-based data augmentation}
\end{figure}

\section{Experiments}
The main goal of the experiments was to examine if GAN-generated data makes classification models more prone to biases. We selected a skin lesion ISIC dataset for distinguishing between malignant and benign lesions.
Our procedure consists of three main steps: data generation, manual artifacts annotation and counterfactual bias insertion. The steps are presented in Fig.~\ref{fig:pipeline}.
For the data generation, we explored unconditional and conditional settings and evaluated their performance in terms of fidelity, diversity and training speed. The generated data was examined in terms of bias inheritance, and further annotated with selected artifacts. We present the statistics and our remarks in the Section~\ref{sec:data_stats}. 
Then, we train our classification models with different data configurations for both unconditional and conditional GANs: classic approach (training on the real data),
augmentation approach (both real and synthetic data), and GANs-only (synthetic data). Each mode is tested how they respond to counterfactual bias insertion. The details behind CBI are presented in the Section~\ref{section:cbi}.

\subsection{Data and training details} 

All our experiments were performed using ISIC Archive challenges 2019 \cite{tschandl2018ham10000,codella2018skin,combalia2019bcn20000} and 2020 \cite{bib:ISIC20} data as our main datasets\footnote{\url{https://www.kaggle.com/nroman/melanoma-external-malignant-256}}.
We splited that dataset randomly into a training set (30~118 samples) and a test set (7~530 samples) both for classification and generation tasks. In some experiments the training subset was augmented with artificial samples, while the test subset remained the same for all conducted studies\footnote{ Data, annotations and additional results are publicly available on GitHub repository: \url{ https://github.com/AgaMiko/debiasing-effect-of-gans}}. Detailed statistics are presented in Supplementary Table~1.

Image generation was performed using the StyleGAN2-ADA modified implementation from NVIDIA Research group\footnote{\url{https://github.com/aidotse/stylegan2-ada-pytorch}}. The ADA mechanism stabilized training in limited data regimes that we faced in malignant samples. To select the best model, we considered both the Fréchet Inception Distance (FID)~\cite{bib:FID} and Kernel Inception Distance (KID)~\cite{bib:KID} metrics, along with training speed, similarly as proposed in~\cite{bib:ganrew}. Achieved results are presented in Supplementary Table~2.


As for the classification model, we used pre-trained EfficientNet-B2~\cite{tan2019efficientnet} and trained it for 20 epochs with an early stopping with three epochs patience. We used Adam optimizer with an adaptive learning rate initialized to 5e-4 and batch size 32. 


\subsection{Descriptive statistics} 
\label{sec:data_stats}
To better understand a skin lesion dataset, or more precisely, the distribution of the artifacts, we have manually annotated 6000 real and synthetic images of skin lesions. We distinguish three main groups with two thousand annotations each: authentic images (real), synthetic images generated with unconditional GANs trained only on images from one class (GAN), and conditional GANs (cGAN). The exact numbers of annotated images are presented in Table~\ref{tab:labeled_stats}.

\begin{figure}\centering
\includegraphics[width=0.8\textwidth]{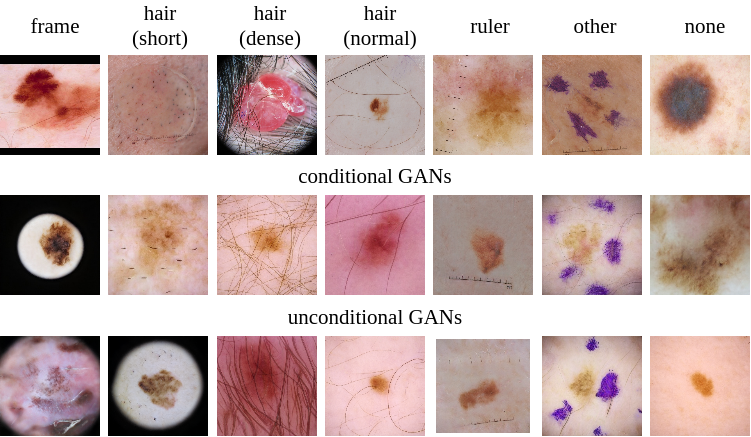}
\caption{Example artifacts in real and GAN generated data.} \label{fig:example_artifacts}
\end{figure}

Based on the literature, we selected four types of artifacts for annotations: hair, frames, rulers and \textit{other} (see Fig.~\ref{fig:example_artifacts}). \textit{Hair} is defined as thick and thin hair of various colors, from light blond to black. Additionally, we annotated  hair types: \textit{normal}, \textit{dense} (covering a significant part of an image) and \textit{short} (shaved hair).
\textit{Frames} are black and white round markings around the skin lesion, black rectangle edges, and vignettes. 
\textit{Rulers} can come in different shapes and colors, either fully or partially visible. 
\textit{Other} are any other artifacts visible that are not as common as ruler marks, frames, and hair. It includes dermatoscopic gel or air bubbles, ink or surgical pen markings, patches and papers, dates and numbers, background parts, light reflection, and dust.

The annotation process was carried out by a trained professional working with the ISIC collection and identification of its biases for over 4 years. Additionally, we measured Inter-Annotator Agreement on a small subsample of data. The mean Cohen’s kappa coefficient was over 70\%, with the highest values on \textit{ruler} annotations and lowest on the \textit{other}.

Interestingly, it seems that both, unconditional and conditional GANs, generate fewer artifacts than in the original images. Most rare or minor artifacts (like dust, paper, number, and dates) are never generated, leading to a significant decrease in the number of images with at least a single artifact. For instance, in unconditional GANs, almost half benign images were rendered without any artifacts. 
Moreover, in GANs, the artifacts are rarely correlated with each other, which means that there is usually one single artifact in the image at a time. The correlation calculated between each artifact and skin lesion type is presented in the Supplementary Table~4. 

There is also a significant reduction in a number of hair and rulers generated in unconditional GANs and a slight one for conditional GANs. Short hair is pretty rare in the original dataset, but they almost entirely vanish in GAN-generated examples. Interestingly, manual annotations showed that conditional GANs seem to generate two rulers in one image of the benign class, which did not happen in the case of malignant skin lesions. This might be connected to the type of rulers annotated, as the GANs almost never generate small, partially visible rulers that are more common in real data.

Similarly, the surgical pen markings were generated only for the benign class in conditional and unconditional GANs, with no single example in the dataset with generated pen marking for the malignant class.
The selectivity in artifacts generation can also be observed in the frame artifact. Frames are a common artifact strongly correlated with the skin lesion category: there are five times more examples of malignant skin lesions with frames than benign. This also affected the GANs training, as in the generated dataset, we observe much more images with frames for both GANs. Even more concerning is the fact that GANs generated only slightly visible vignettes or tiny, few pixel rectangular frames for benign moles. There was no single case of benign skin lesion generated with a large black frame, round frame, or strong vignette. All frames in benign class were almost invisible. On the contrary, the malignant class was always present with large round black frames or strong vignettes. This alone shows a huge amplifying effect on already pre-existing solid bias in the dataset.

This concludes that GANs might amplify strong biases but mitigate the small ones making it a double-edged sword. GANs might increase already strong biases (or essential features), leading to even more spurious correlations and, at the same time, loose information about insignificant tendencies and rare patterns. This property might be connected to the GANs architectures (such as kernel filter sizes) or the number of artifacts in the training dataset.  Additionally, we provide the Predictive Power Scores (PPS) that, in contrast, to feature importance, are calculated using only a single feature (here: an artifact) trying to predict the target column (malignant/benign class)~\cite{florian_wetschoreck_2020_4091345}. The result supports our conclusion about (de)biasing effect of GAN-based augmentation. The scores are presented in Supplementary Table~\ref{tab:pps}.


\begin{table}[]
\caption{Statistics for manually labeled real and artificial images of malignant (mal) and benign (ben) class. cGAN refers to conditional GAN, while GAN -- unconditional trained only on images from one class.}
\label{tab:labeled_stats}
\resizebox{\textwidth}{!}{%
\begin{tabular}{|l|l|c|c|c|c|c|c|c|c|}
\hline
\textbf{} & class & \multicolumn{1}{l}{hair (normal)} & \multicolumn{1}{l}{hair (dense)} & \multicolumn{1}{l}{hair (short)} & \multicolumn{1}{l}{ruler} & \multicolumn{1}{l}{frame} & \multicolumn{1}{l}{other} & \multicolumn{1}{l}{none} & \multicolumn{1}{l|}{\textbf{total}} \\ \hline
\textbf{Real} & ben & 467 & 110 & 45 & 211 & 57 & 201 & 269 & 1000 \\
\textbf{} & mal & 444 & 50 & 51 & 287 & 251 & 402 & 141 & 1000 \\
\textbf{cGAN} & ben & 319 & 57 & 8 & 186 & 84 & 106 & 352 & 1000 \\
\textbf{} & mal & 223 & 29 & 8 & 110 & 365 & 128 & 328 & 1000 \\
\textbf{GAN} & ben & 190 & 43 & 4 & 94 & 78 & 257 & 412 & 1000 \\
 & mal & 234 & 40 & 16 & 41 & 381 & 197 & 289 & 1000\\
 \hline
\end{tabular}%
}
\end{table}

\subsection{Counterfactual bias insertion} \label{section:cbi} 
The previous section identified three possible sources of bias in skin lesion classification: hair (regular, short, and dense), black frames, and ruler marks. 
We have tested several different ways and proportions of real to synthetic data to find the best performance metrics, as we wanted to mimic the realistic approach to data augmentation.  We achieved the best scores when augmenting only the malignant class with 15k synthetic images.
Achieved results are described in Table~\ref{tab:counterfactual}.

We use the CBI metrics~\cite{mikolajczyk2020towards} to measure bias influence. 
Frame bias insertion was done by adding a frame to an image. Hair and ruler insertion required more care to achieve a realistic look. We copied artifacts from the source image to the target image using provided segmentation masks~\cite{ramella2021hair}. 
We selected samples for each bias for a broader analysis, resulting in five frames, five types per hair type (regular, short, and dense), and five rulers. The segmentation masks used for the analysis are provided in Supplementary Fig.~1. The CBI was calculated for each image in the dataset, by inserting each of 25 biases. The mean CBI scores for each bias group are presented in Table~\ref{tab:counterfactual}.  As the results strongly depended on the segmentation mask used, we also calculated standard deviation per bias.

\begin{table}[!htb]
\caption{CBI metrics and $F_1$ scores measuring bias influence for each selected bias and type of training data: real data (real), augmented with synthetic images (aug.) and synthetic data (synth.), generated both with conditional (cGAN) and unconditional GANs (GAN) trained only on one class: benign (ben) or malignant (mal). Higher $F_1$ score means better performance, while higher number of switched images mean a higher bias influence.}
\resizebox{\textwidth}{!}{%
\begin{tabular}{|l|l|rrrrr|rrrr|}
\hline
\textbf{bias} & \textbf{data} & \multicolumn{5}{c|}{\textbf{switched}} & \multicolumn{4}{c|}{\textbf{$F_1$ (\%)}} \\
\textbf{} & \textbf{} & mean & std$^1$ & median & mal to ben & ben to mal & \textbf{$F_1$} & aug & std$^2$ & mean \\ \hline
\textbf{frame} & real & 129 & 119.39 & 77 & 24 (2.39\%) & 104 (1.60\%) & 91.99 & 88.97 & 4.01 & 90.48 \\
 & aug. cGAN & 223 & 55.25 & 199 & 40 (3.88\%) & 183 (2.81\%) & 89.65 & 84.93 & 2.26 & 87.29 \\
\textbf{} & \textbf{aug. GAN} & \textbf{59} & 16.07 & \textbf{51} & \textbf{22} (2.19\%) & \textbf{37} (0.57\%) & \textbf{91.52} & \textbf{90.49} & 0.61 & \textbf{91.01} \\
 & synth. cGAN & 290 & 43.97 & 271 & 125 (12.24\%) & 165 (2.54\%) & 80.39 & 79.28 & 1.26 & 79.84 \\
 & synth. GAN & 413 & 33.17 & 404 & 297 (29.13\%) & 116 (1.78\%) & 76.04 & 74.99 & 0.82 & 75.51 \\ 
\hline
\textbf{ruler} & \textbf{real} & \textbf{81} & \textbf{86.76} & \textbf{29} & \textbf{76} (7.48\%) & \textbf{5} (0.07\%) & \textbf{91.99} & \textbf{88.59} & \textbf{4.30} & \textbf{90.29} \\
 & aug. cGAN & 79 & 44.21 & 69 & 55 (5.43\%) & 24 (0.37\%) & 89.65 & 89.18 & 1.08 & 89.41 \\
 & aug. GAN & 81 & 96.08 & 24 & 78 (7.60\%) & 3 (0.05\%) & 91.52 & 87.05 & 5.81 & 89.29 \\
 & synth. cGAN & 200 & 137.26 & 151 & 194 (18.96\%) & 6 (0.09\%) & 80.39 & 78.31 & 5.11 & 79.35 \\
 & synth. GAN & 154 & 109.89 & 107 & 65 (6.33\%) & 90 (1.38\%) & 76.04 & 74.69 & 1.82 & 75.36 \\ 
\hline
\textbf{dense} & \textbf{real} & \textbf{109} & \textbf{33.63} & \textbf{118} & \textbf{90} (8.81\%) & \textbf{19} (0.29\%) & \textbf{91.99} & \textbf{88.42} & \textbf{1.62} & \textbf{90.20} \\
 & aug. cGAN & 439 & 269.40 & 459 & 96 (9.38\%) & 344 (5.28\%) & 89.65 & 78.85 & 9.04 & 84.25 \\
 & aug. GAN & 122 & 28.48 & 113 & 74 (7.29\%) & 48 (0.73\%) & 91.52 & 87.03 & 1.42 & 89.28 \\
 & synth. cGAN & 325 & 71.38 & 357 & 272 (26.66\%) & 52 (0.81\%) & 80.39 & 80.00 & 1.43 & 80.20 \\
 & synth. GAN & 1089 & 651.43 & 1101 & 61 (5.97\%) & 1028 (15.79\%) & 76.04 & 59.94 & 10.27 & 67.99 \\ 
\hline
\textbf{medium} & \textbf{real} & \textbf{27} & \textbf{7.37} & \textbf{26} & \textbf{17} (1.63\%) & \textbf{10} (0.15\%) & \textbf{91.99} & \textbf{91.60} & \textbf{0.14} & \textbf{91.79} \\
 & aug. cGAN & 74 & 17.85 & 74 & 38 (3.74\%) & 36 (0.55\%) & 89.65 & 89.31 & 0.97 & 89.48 \\
 & aug. GAN & 28 & 8.23 & 26 & 12 (1.19\%) & 16 (0.25\%) & 91.52 & 91.11 & 0.25 & 91.32 \\
 & synth. cGAN & 163 & 47.93 & 177 & 113 (11.05\%) & 50 (0.77\%) & 80.39 & 80.49 & 1.84 & 80.44 \\
 & synth. GAN & 284 & 141.58 & 298 & 46 (4.47\%) & 238 (3.66\%) & 76.04 & 73.51 & 3.20 & 74.78 \\ 
\hline
\textbf{short} & real & 77 & 99.49 & 38 & 67 (6.52\%) & 10 (0.16\%) & 91.99 & 88.72 & 5.21 & 90.35 \\
 & aug. cGAN & 180 & 114.84 & 224 & 12 (1.16\%) & 168 (2.59\%) & 89.65 & 84.73 & 3.56 & 87.19 \\
 & \textbf{aug. GAN} & \textbf{54} & \textbf{50.91} & \textbf{32} & \textbf{37} (3.64\%) & \textbf{17} (0.26\%) & \textbf{91.52} & \textbf{89.55} & \textbf{2.40} & \textbf{90.54} \\
 & synth. cGAN & 249 & 135.44 & 282 & 221 (21.67\%) & 28 (0.43\%) & 80.39 & 78.80 & 1.31 & 79.60 \\
 & synth. GAN & 380 & 445.91 & 191 & 57 (5.62\%) & 323 (4.96\%) & 76.04 & 70.36 & 9.30 & 73.20 \\
 \hline
\end{tabular}%
} \small 
$^1$ -- standard deviation for \textit{switched} metric for different bias types, \\
$^2$ -- standard deviation for $F_1^{aug}$. STD for $F_1$ is equal to 0.
\label{tab:counterfactual}
\end{table}

Experiments allow understanding of how each artifact type affects the training, e.g., thin frames usually make predictions switch from malignant to benign, and large frames from benign to malignant. Rulers usually make predictions shift from malignant to benign, but in the GAN-augmented case, a thin ruler in the bottom causes prediction switch from benign to malignant.

The best performance and CBI scores were for real and augmented (aug. GAN) data. We also analyzed different augmentation policies and found that not every approach with augmentation gives better results than real data. Only the proposed approach did not provide worse CBI results than real. In all cases, the worst scores were observed for synthetic datasets. In general higher $F_1$ scores seemed to be a surprisingly accurate measure in case of vulnerability to biases. However, it also appears that quite a high score ($\>90\%$) is needed to trust it: models with lower $F_1$ were not necessarily less biased.

Additionally, it is worth noticing that unconditional GANs performed better and were less prone to learn biases. Better performance might be connected with the lower Perceptual Path Length (PPL)~\cite{bib:ppl} scores in unconditional GANs (see Supplementary Table~2). PPL measures the difference between consecutive images when interpolating between two random inputs. Lower PPL scores mean that the latent space is regularized better. Here, unconditional GANs have to learn the pattern distribution of only one class: either malignant or benign. We hypothesized this is also one of the reasons why unconditional GANs are better at capturing the consistency of the images. In contrast, cGANs seemed to link some biases to a one, specific class, resulting in a more biased dataset.



\section{Conclusions}

Descriptive statistics indicated that GANs amplified strong biases: large black frames, common dermoscopy artifacts, were never generated in benign skin lesions but were more prevalent in the generated dataset than in the original one. At the same time, the amount of clean images was much higher in the case of synthetic images. This observation and the manual exploration of generated artifacts implied that GANs also have debiasing properties, especially in the case of small, rare biases. 
In addition, for better reproducibility of our studies we provided manual annotations of biasing artifacts, which can serve as a benchmark for further studies. Future directions will be focused on generating unbiased data by learning directions for each of the biases in the latent space, to create a more complex, fair and diverse dataset. 

The counterfactual bias insertion analysis supported the theory of GANs (de)biasing attributes. The study demonstrated an inverted correlation between the model's accuracy and bias robustness. This suggested that a well-trained model, even on biased data, is less likely to switch predictions after inserting biases. Ultimately, the best results in terms of accuracy and robustness were achieved for models trained on real data, or augmented with synthetic images produced by unconditional GANs. This shows that GANs can be successfully used to enrich data but should be monitored, as they can amplify preexisting inequities in data.


\section{Acknowledgements}

The  research on bias reported  in  this  publication  was supported  by  Polish  National  Science  Centre (Grant Preludium No: \textit{UMO-2019/35/N/ST6/04052}). GANs trainings were conducted during first rotations of the \textit{Eye for AI Program} thanks to the support of Sahlgrenska University Hospital and AI Sweden. 
\bibliographystyle{splncs04}
\bibliography{mybibliography}

\clearpage
\setcounter{secnumdepth}{0}
\setcounter{figure}{0} 
\setcounter{table}{0}
\setcounter{page}{1}
\section{Supplementary materials}

\begin{table}[]
\centering
\caption{Exact number of real and synthetic images for each proposed experiment. In case of GANs training we used images only from training subset.}
\label{tab:train_stats}
\begin{tabular}{|l|ll|ll|ll|l|}
\hline
 & \multicolumn{2}{c|}{\textbf{real}} & \multicolumn{2}{c|}{\textbf{synthetic}} & \multicolumn{2}{c|}{\textbf{total}} & \multicolumn{1}{c|}{\textbf{mode}} \\
 & ben & mal & ben & mal & ben & mal &  \\ \hline
\textbf{train} & 26033 & 4085 & - & - & 26033 & 4085 & real \\
 & 26033 & 4085 & - & 15000 & 26033 & 19085 & augmented \\
 & - & - & 27500 & 27500 & 27500 & 27500 & synthetic \\ \hline
\textbf{test} & 6509 & 1021 & - & - & 6509 & 1021 & real, augmented, synthetic \\
\hline
\end{tabular}
\end{table}

\begin{table}
\centering
\caption{Calculated metrics for each of the tested generative models. The best KID and FID scores were achieved for conditional StylGAN2-ADA without color augmentations. Unconditional GAN for non-melanoma is slightly better in terms of precision and recall. The unconditional models have lower PPL scores, showing a better regularization of the latent space due to the fact that here we modeled only the distribution of one class.}
\begin{tabular}{|ll|rrrrr|}
\hline
     & class               & KID (\%)       & FID           & Precision     & Recall         & PPL \\
\hline
GAN  & mal                 & 0.42           & 7.99          & \textbf{0.77} & \textbf{0.45}  & 60 \\
GAN  & ben                 & 0.47           & 15.46         & 0.62          & 0.40           & \textbf{51} \\
cGAN & ben and mal         & \textbf{0.24}  & \textbf{7.02} & 0.75          & 0.44           & 101 \\
\hline
\end{tabular}
\label{tab:gans_metrics}
\end{table}

\begin{figure}
\centering
\includegraphics[width=0.6\textwidth]{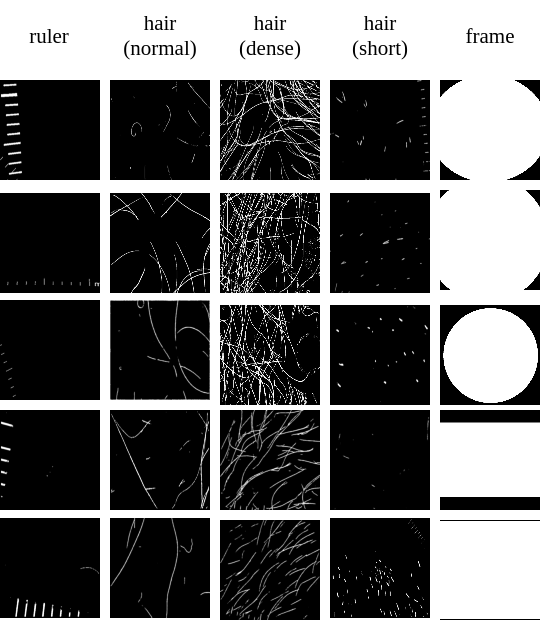}
\caption{Segmentation masks used to insert artifacts with Counterfactual Bias Insertion method.} 
\label{fig:masks_artifacts}
\end{figure}

\begin{table}[]
\centering
\caption{PPS score calculated for every feature with metric weighted F1 and a Decision Tree Classifier. PPS score was calculated for every combination of artifacts with type, and only the ones above 0 are presented.}
\begin{tabular}{|l|llllll|}
\hline
\textbf{feature} & \textbf{value} & \textbf{PPS real} & \textbf{GANc} & \textbf{GANu} & \textbf{baseline\_F1} & \textbf{model\_F1} \\ \hline
hair & type & 0.55\% &  & 9.40\% & 51.86\% & 52.13\% \\
frame & type & 6.76\% & 23.87\% & 22.18\% & 51.86\% & 55.12\% \\
other & type & 13.40\% &  &  & 51.86\% & 58.31\% \\ \hline
\end{tabular}
\label{tab:pps}
\end{table}

\begin{table}[]
\caption{Correlation matrix. Artifacts and class correlation is calculated for unconditional GAN, conditional GAN (cGAN), and real images.}
\centering
\begin{tabular}{|ll|lllll|}
\hline
 &  & \textbf{hair} & \textbf{frame} & \textbf{ruler} & \textbf{other} & \textbf{type} \\ \hline
GAN & \textbf{hair} &  & -0.26\% & -3.43\% & -15.61\% & 6.02\% \\
cGAN & \textbf{} &  & -3.48\% & -13.65\% & -3.45\% & -13.27\% \\
Real & \textbf{} &  & -4.46\% & -8.79\% & -12.11\% & -4.49\% \\ \hline
GAN & \textbf{frame} & -0.26\% &  & -13.26\% & -9.99\% & 36.03\% \\
cGAN & \textbf{} & -3.48\% &  & -19.73\% & -3.18\% & 33.67\% \\
Real & \textbf{} & -4.46\% &  & 8.65\% & -4.21\% & 28.72\% \\ \hline
GAN & \textbf{ruler} & -3.43\% & -13.26\% &  & -7.44\% & -10.56\% \\
cGAN & \textbf{} & -13.65\% & -19.73\% &  & 0.16\% & -10.70\% \\
Real & \textbf{} & -8.79\% & 8.65\% &  & 0.94\% & 9.46\% \\ \hline
GAN & \textbf{other} & -15.61\% & -9.99\% & -7.44\% &  & -7.16\% \\
cGAN & \textbf{} & -3.45\% & -3.18\% & 0.16\% &  & 3.42\% \\
Real & \textbf{} & -12.11\% & -4.21\% & 0.94\% &  & 21.18\% \\ \hline
GAN & \textbf{class (mal)} & 6.02\% & 36.03\% & -10.56\% & -7.16\% &  \\
cGAN & \textbf{} & -13.27\% & 33.67\% & -10.70\% & 3.42\% &  \\
Real & \textbf{} & -4.49\% & 28.72\% & 9.46\% & 21.18\% &  \\ \hline
\end{tabular}

\label{tab:my-table}
\end{table}

\end{document}